\documentclass[conference]{IEEEtran}
\IEEEoverridecommandlockouts
\usepackage{cite}
\usepackage{amsmath,amssymb,amsfonts}
\usepackage{algorithmic}
\usepackage{graphicx}
\usepackage{textcomp}
\usepackage{xcolor}
\usepackage{multirow}
\usepackage{url}
\begin{document}

\title{Real-time Neuron Segmentation for Voltage Imaging
}

\newcommand{\markchanges}[1]{#1}

\newcommand{\fixstars}{Fixstars Solutions, Inc.}

\author{
\IEEEauthorblockN{Yosuke Bando}
\IEEEauthorblockA{Kioxia Corporation}
\and
\IEEEauthorblockN{Ramdas Pillai}
\IEEEauthorblockA{\fixstars}
\and
\IEEEauthorblockN{Atsushi Kajita}
\IEEEauthorblockA{\fixstars}
\and
\IEEEauthorblockN{Farhan Abdul Hakeem}
\IEEEauthorblockA{\fixstars}
\and
\IEEEauthorblockN{Yves Quemener}
\IEEEauthorblockA{\fixstars}
\and
\IEEEauthorblockN{Hua-an Tseng}
\IEEEauthorblockA{Boston University}
\and
\IEEEauthorblockN{Kiryl D. Piatkevich}
\IEEEauthorblockA{Westlake University}
\and
\IEEEauthorblockN{Changyang Linghu}
\IEEEauthorblockA{University of Michigan}
\and
\IEEEauthorblockN{Xue Han}
\IEEEauthorblockA{Boston University}
\and
\IEEEauthorblockN{Edward S. Boyden}
\IEEEauthorblockA{MIT \& \\ Howard Hughes Medical Institute}
}

\IEEEpubid{\begin{minipage}{.9\textwidth}\
    \\[15mm]
    IEEE International Conference on Bioinformatics and Biomedicine (BIBM), 813--818, 2023.\\
    \url{https://ieeexplore.ieee.org/document/10385929}\\
    \copyright~2023 IEEE. Personal use of this material is permitted. Permission from IEEE must be obtained for all other uses, in any current or future media, including reprinting/republishing this material for advertising or promotional purposes, creating new collective works, for resale or redistribution to servers or lists, or reuse of any copyrighted component of this work in other works.
\end{minipage}}

\maketitle

\begin{abstract}
In voltage imaging, where the membrane potentials of individual neurons are recorded at from hundreds to thousand frames per second using fluorescence microscopy, data processing presents a challenge.
Even a fraction of a minute of recording with a limited image size yields gigabytes of video data consisting of tens of thousands of frames, which can be time-consuming to process.
Moreover, millisecond-level short exposures lead to noisy video frames, obscuring neuron footprints especially in deep-brain samples where noisy signals are buried in background fluorescence.
To address this challenge, we propose a fast neuron segmentation method able to detect multiple, potentially overlapping, spiking neurons from noisy video frames, and implement a data processing pipeline incorporating the proposed segmentation method along with GPU-accelerated motion correction.
By testing on existing datasets as well as on new datasets we introduce, we show that our pipeline extracts neuron footprints that agree well with human annotation even from cluttered datasets, and demonstrate real-time processing of voltage imaging data on a single desktop computer for the first time.
\end{abstract}


\section{Introduction}
\label{sec:intro}

\begin{figure*}[t]
\centering
\includegraphics[trim=0 290 0 0,clip,width=\linewidth]{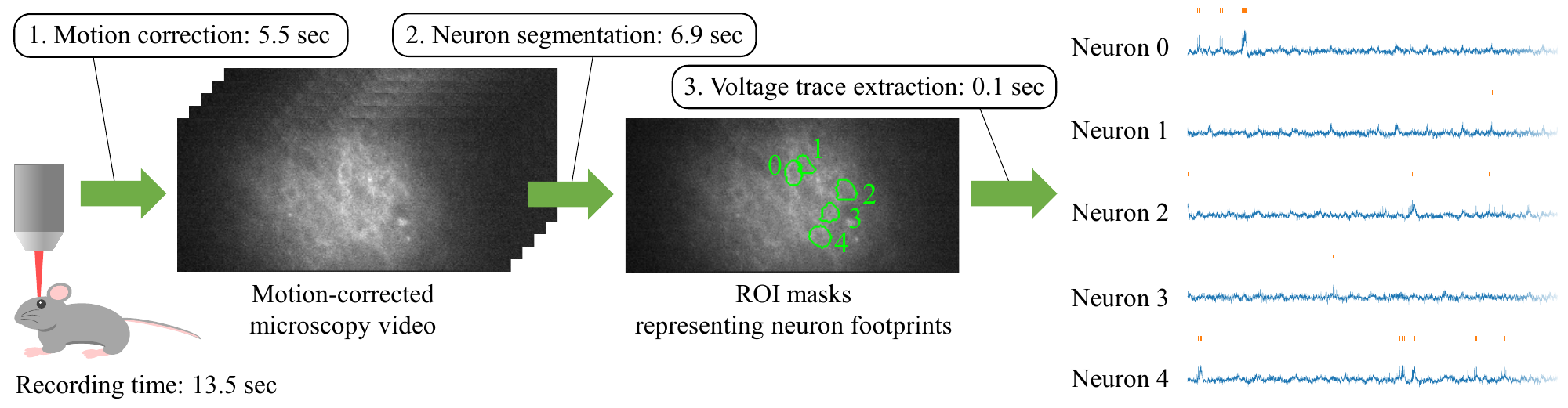}
\caption{Voltage imaging data processing pipeline.
Our pipeline processes an input video in real time, meaning that the processing time is shorter than the recording time.
As an example, given a video recorded in 13.5 sec capturing 10,000 frames at 741 fps, the three stages of the pipeline spend 5.5 sec, 6.9 sec, and 0.1 sec, respectively, totaling 12.5 sec, which is shorter than the recording time of 13.5 sec.}
\label{fig:pipeline}
\end{figure*}

Voltage imaging uses fluorescence microscopy to monitor neural activities of animals \cite{VoltageImaging}.
It uses fluorophores called {\it voltage indicators} that change their fluorescence depending on membrane potentials of neurons, providing signals that are close to neural voltage measured using more invasive, physically contacting devices such as patch clamps and electrodes.
By capturing images at from hundreds to thousand frames per second (fps), voltage imaging enables temporally high-resolution detection of spikes as well as extraction of subthreshold activities, which is an advantage over more established Calcium imaging that uses Calcium ion concentration as a 
slow proxy for rapidly changing membrane potential \cite{VoltageVsCa}.

However, voltage imaging presents a challenge in data processing.
Even a fraction of a minute of recording with a limited image size 
yields gigabytes of video data consisting of tens of thousands of frames, and processing it to extract voltage traces from captured neurons can be time-consuming.
Existing methods either require manual annotation \cite{Voltron,SomArchon,Simul2pVoltage,HighSpeed2pVoltage} or spend significantly more time on image processing than the image recording time \cite{InVivo,VolPy}, both of which can slow down iterative experiments by neuroscientists.
Moreover, video frames captured with millisecond-level short exposures have a low signal-to-noise (SNR) ratio despite significant improvements in brightness and sensitivity of voltage indicators in recent years.
Although the SNR of a voltage trace can be improved by combining observations from multiple pixels belonging to the same neuron, it can be challenging to identify (either manually or automatically) where neurons are in the video in the first place, if noisy signals are buried in background fluorescence especially in deep-brain samples.
While background fluorescence may be reduced by two-photon microscopy \cite{Ultra2p,Simul2pVoltage,HighSpeed2pVoltage} or optical techniques that steer light onto neurons of interest \cite{paQuasAr,VoltageDMD}, it is desirable to be able to use more prevalent, unmodified one-photon microscopes.

This paper presents a data processing pipeline that can segment footprints of spiking neurons from noisy voltage imaging data in real time, meaning that the runtime is equal to or shorter than the video recording time, on a single desktop computer.
%
We build on the idea from previous work \cite{VolPy} that summarizes a video into a few still images.
However, since we find there are cases where it is challenging to identify neurons given a few still images alone, 
our proposal is to split a video into time segments, and apply a summary image approach to them individually.
We use the U-Net convolutional neural network (CNN) \cite{UNet} to identify spiking neurons from summary images for each time segment, and aggregate them into a single set of ROI masks.
This better exploits temporal information while still benefiting from reduced computation by summarization.
In combination with GPU-accelerated motion correction we develop as a step before the segmentation, our pipeline as a whole runs in real time while still leaving some room for voltage trace extraction from ROIs.

\IEEEpubidadjcol

We run our pipeline on existing datasets as well as on new datasets we introduce, and show that the processing times for all of the datasets are shorter than the respective video recording times.
We also show that the ROI masks produced by our method have around 80\% agreement (more precisely, an $F_1$ score of 0.8) with manual annotation on average. 

In summary, the contributions of this paper are as follows.
\begin{itemize}
\item We propose a fast neuron segmentation method able to detect multiple, potentially overlapping, spiking neurons whose extracted footprints agree well with human annotation even for cluttered one-photon datasets.
\item We implement a data processing pipeline incorporating the proposed segmentation method along with GPU-accelerated motion correction, and demonstrate real-time processing of voltage imaging data on a single desktop computer for the first time.
\end{itemize}


\section{Pipeline Overview}

Our pipeline consists of three stages as shown in Figure~\ref{fig:pipeline}.
\begin{enumerate}
\item {\bf Motion correction} for canceling motion so that the corrected video shows stationary neurons whose intensity variation comes from their changes in fluorescence.
\item {\bf Neuron segmentation} for detecting neurons and delineating their contours from the background.
\item {\bf Voltage trace extraction} to estimate the time-varying membrane potential of each segmented neuron.
\end{enumerate}


The primary focus of this paper is the second stage of the pipeline: we propose a fast method for segmenting spiking neurons, which will be explained in Section~\ref{sec:segment}.
%

Here we briefly describe the other two stages.
Our motion correction consists of carefully engineered implementations of well-known techniques.
It calculates the zero-mean normalized cross-correlation (ZNCC) between images to measure their similarity.
We tile 21x21-pixel patches to cover the image, and compute ZNCC for all the patches and candidate motion vectors in parallel on the GPU, while employing optimizations via table-based area sums \cite{FastTemplateMatching}.
Once patch-wise ZNCC values are computed, the GPU threads are synchronized and the values are aggregated to identify the most likely motion vector.
%
Our voltage trace extraction takes the mean pixel intensity within each detected neuron ROI for each frame of the motion-corrected video.
This is a rudimentary method for reference: more sophisticated alternatives may be used \cite{Voltron,paQuasAr,InVivo,VolPy}.

\section{Neuron Segmentation}
\label{sec:segment}

This section presents details of the segmentation stage.

\subsection{Motivation and Design}
\label{sec:motivation}

The previous methods for segmenting neurons from voltage imaging data take quite different approaches from each other.
SGPMD-NMF \cite{InVivo} applies local nonnegative matrix factorization (NMF) to a video to decompose it into components each having spatially contiguous, temporally correlated pixels.
Those components are considered neuron footprints.
While this method has been shown to extract voltage signals including subthreshold dynamics with high fidelity, it takes time:
using our desktop computer, it takes 33 min to analyze a 450x138-pixel video of 10,000 frames including the denoising time necessary before the local NMF (a similar runtime is reported in \cite{InVivo} on a computing cluster).
Since the recording time of this video is 13.5 sec, the processing time is 148 times longer.
SGPMD-NMF also requires some human intervention to identify blood vessels and initialize background regions.

In contrast, VolPy \cite{VolPy} is relatively fast and fully automatic.
It summarizes an entire video into a few still images, which are input to the Mask R-CNN \cite{MaskRCNN} 
in order to obtain neuron ROI masks.
On top of the fact that it has to deal with only a few images, a CNN-based method is fast once trained, which makes a summary image method an appealing approach in terms of speed.
While there are usually trade-offs between accuracy and speed, in this paper we are more interested in a speed-oriented solution that allows neuroscientists to iterate experiments quickly.
Once good initial results are obtained, more elaborate analysis like SGPMD-NMF may be used later.

\begin{figure}[t]
    \centering
    \includegraphics[width=0.4\columnwidth]{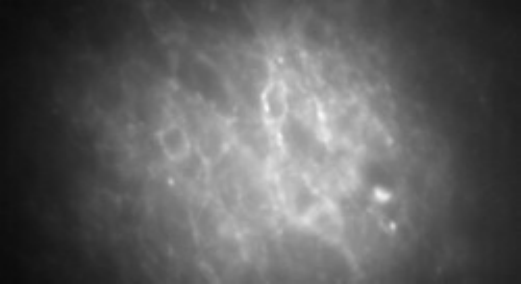}
    \includegraphics[width=0.4\columnwidth]{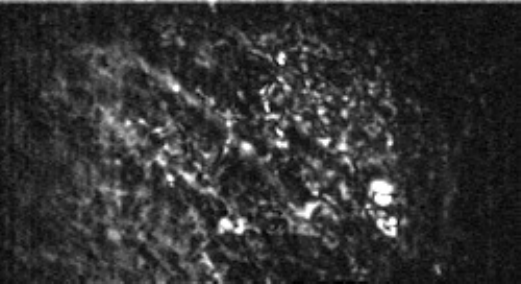}\\
    \mbox{\footnotesize \hspace{1mm} (a) Temporal average \hspace{8mm} (b) Temporal correlation}
    \caption{Summary images used in VolPy \cite{VolPy} for the input video in Figure~\ref{fig:pipeline}.}
    \label{fig:summary_images}
\end{figure}

\begin{figure*}[t]
    \centering
    \includegraphics[trim=0 208 0 0,clip,width=0.9\linewidth]{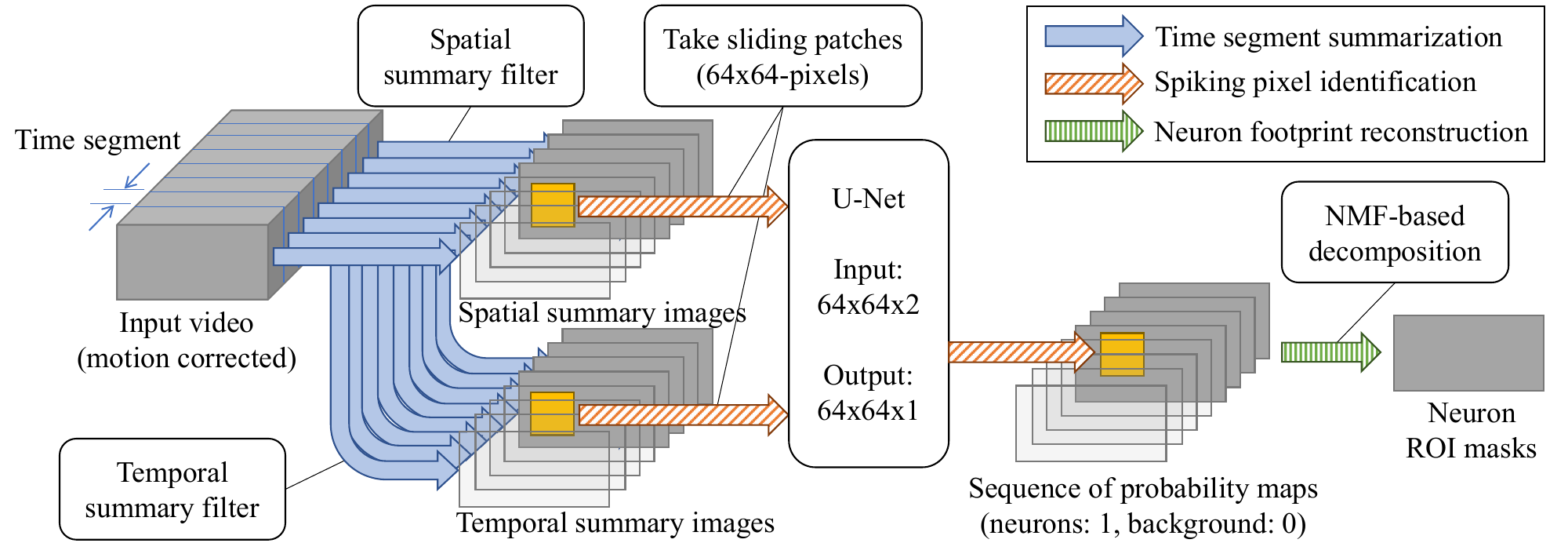}
    \caption{Proposed segmentation subpipeline. The depth direction represents the time axis.}
    \label{fig:segmentation}
\end{figure*}

That being said, in order to deal with noisy one-photon voltage imaging data, we find summary image methods can be challenging.
As an example, Figure~\ref{fig:summary_images} shows the temporal average and correlation images used by VolPy computed for the input video shown in Figure~\ref{fig:pipeline}, which will be subsequently input to the Mask R-CNN.
As can be seen, although the temporal average image provides much less noisy clues to neuron boundaries, the cluttered background makes it hard to delineate them.
The temporal correlation image unfortunately provides little information on where neurons are either, because the correlation between nearby pixels is buried in noise.

Thus, building on these previous works, we take a middle-ground approach where a video is split into time segments which are individually summarized and processed by a CNN
as shown in Figure~\ref{fig:segmentation}.
This gives us a sequence of probability maps representing where neurons are likely to be spiking during each time segment, which is subsequently aggregated into ROI masks via NMF.
This better exploits temporal information than immediately reducing the entire video into a few images, while still reducing computation by summarization.

\subsection{Time Segment Summarization}


We use two summary filters that each project a time segment along the time axis to produce a single summary image.
In what follows, we denote the $i$-th time segment of the motion-corrected input video by $V_i(x, t)$, where $x$ represents 2D spatial coordinates and $t \in [1, L]$ represents a frame number within the segment. 
We use a time segment length of $L = 50$ frames throughout this paper, which consistently produced good results for different datasets with varying frame rates.

The first filter is temporal average as
\begin{equation}
\label{eqn:spatial}
S_i(x) = \frac{1}{L} \sum_{t=1}^L V_i(x, t),
\end{equation}
which produces a similar image to Figure~\ref{fig:summary_images}(a), albeit with slightly increased noise due to a smaller number 
of frames to average.
As it reveals spatial image features (i.e., boundaries) of neurons more clearly, we call it a {\it spatial summary image}.

\begin{figure}[t]
\centering
\includegraphics[trim=10 290 390 90,clip,width=0.8\columnwidth]{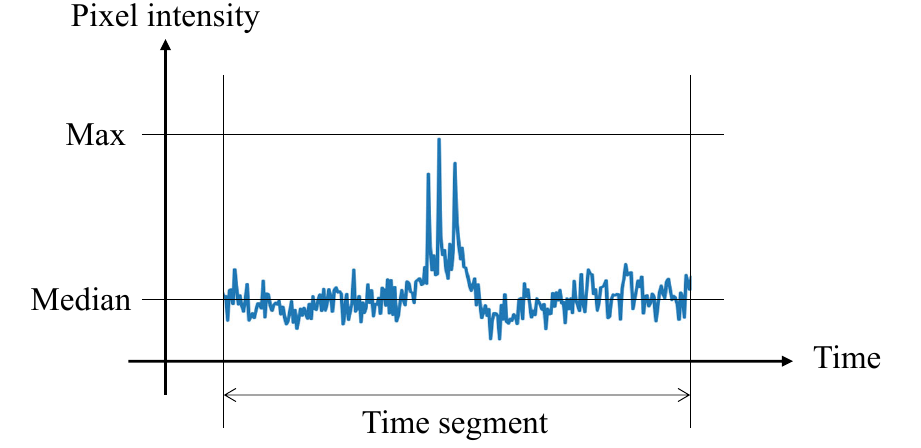}
\caption{Max-median filter.}
\label{fig:max-median}
\vspace{5mm}
\centering
\includegraphics[trim=0 430 570 0,clip,width=0.8\columnwidth]{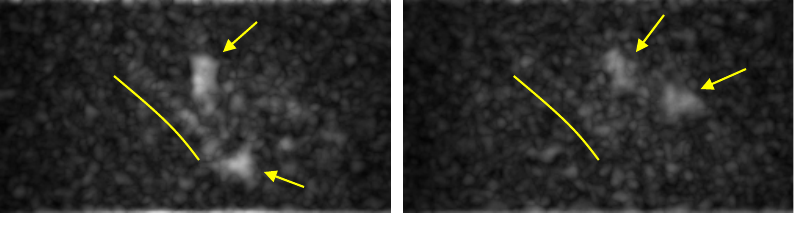}
\caption{Examples of temporal summary images using max-median filter.}
\label{fig:max-median_images}
\end{figure}

The second filter is temporal maximum-minus-median (max-median for short) to enhance spiking neurons as
\begin{equation}
\label{eqn:max_med}
T_i(x) = \max_{t \in [1, L]} \{ \bar{V}_i(x, t) \} - \underset{t \in [1, L]}{\rm median} \{ \bar{V}_i(x, t) \},
\end{equation}
where $\bar{V}_i$ is a spatially smoothed version of $V_i$ using a Gaussian filter (we use a standard deviation of 3 pixels).
If there is a spiking neuron at pixel $x$ during this time segment, the temporal max operator identifies the peak of the spike while the temporal median extracts the baseline potential as shown in Figure~\ref{fig:max-median}, and therefore the difference gives a positive response.
We find that pre-smoothing $\bar{V}_i$ is necessary to have a clear response.
Figure~\ref{fig:max-median_images} shows examples of the filtering results from two time segments of the input video shown in Figure~\ref{fig:pipeline}, each showing two bright blobs (pointed to by the arrows) likely coming from spiking neurons.
They additionally show some linear structures (the most visible one is indicated by the line next to it) from blood vessels as well as small blobs everywhere due to noise.
Since these images reveal locations of temporal activities, we call them {\it temporal summary images}.


\subsection{Spiking Pixel Identification}

\begin{figure}[t]
    \centering
    \includegraphics[trim=15 135 15 20,clip,width=\columnwidth]{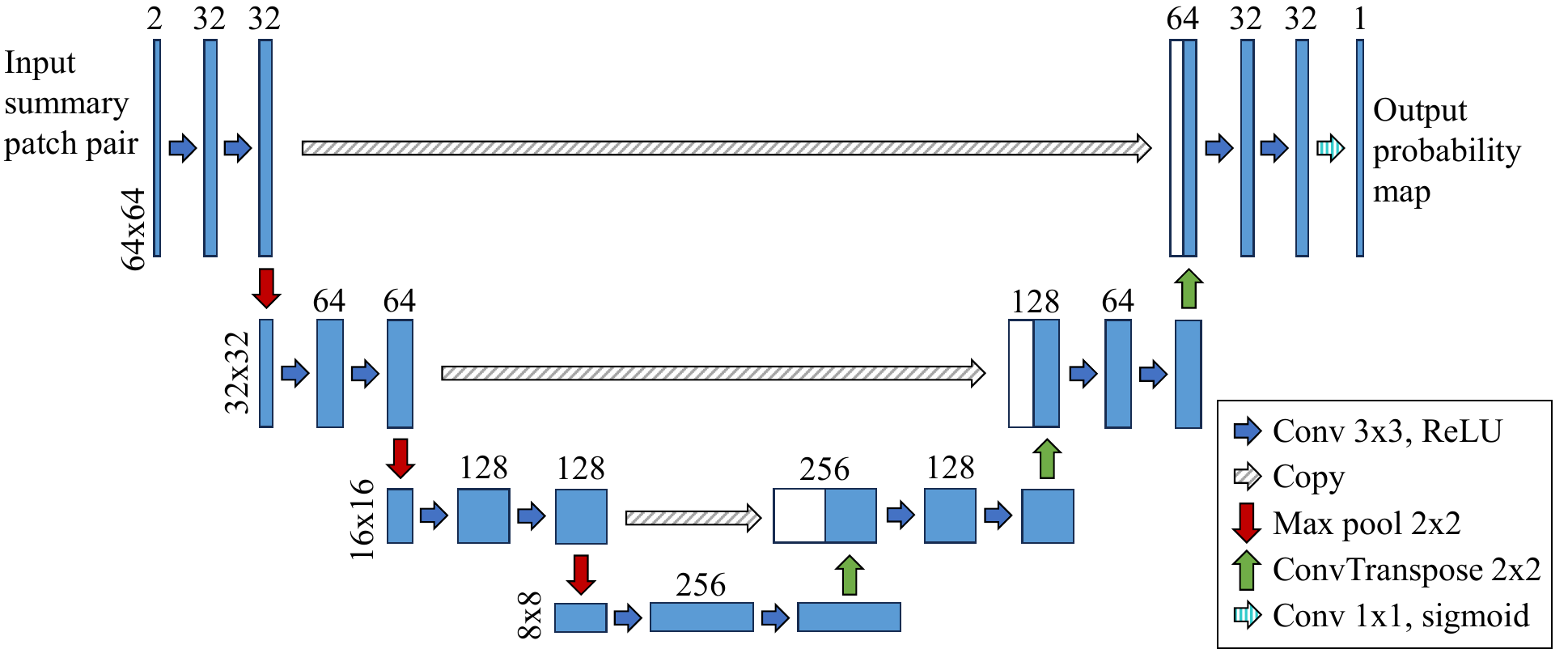}
    \caption{Our lightweight U-Net configuration with a small input size of 64x64. The diagram convention follows the original U-Net work \cite{UNet}.}
    \label{fig:u-net}
\vspace{5mm}
    \centering
    \includegraphics[width=0.2\columnwidth]{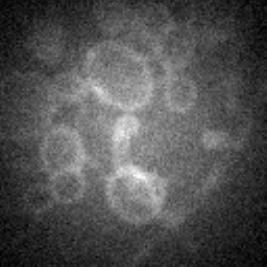}
    \includegraphics[width=0.2\columnwidth]{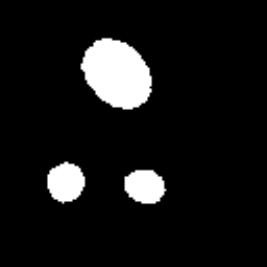}\\
    \caption{Synthetic training data. An example video frame (left) and the corresponding binary mask indicating the footprints of spiking neurons (right).}
    \label{fig:train}
\vspace{5mm}
    \centering
    \includegraphics[width=0.4\columnwidth]{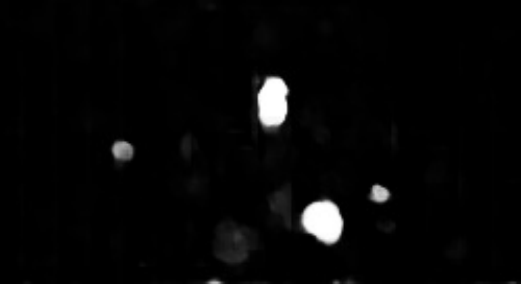}
    \includegraphics[width=0.4\columnwidth]{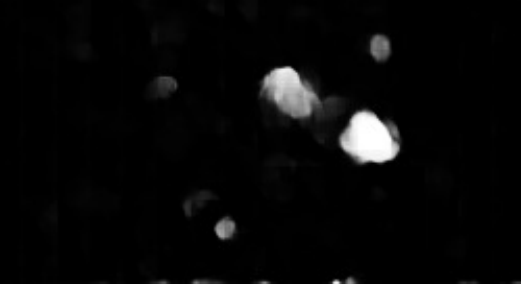}\\
    \caption{U-Net outputs representing the probability of each pixel belonging to spiking neurons.
    The input temporal summary images are shown in Figure~\ref{fig:max-median_images}.}
    \label{fig:u-net_outputs}
\end{figure}

Based on the two summary images $S_i(x)$ and $T_i(x)$, 
we estimate where spiking neurons are likely to be during this ($i$-th) time segment.
To this end, we employ the U-Net, a widely-used CNN for biomedical image segmentation \cite{UNet}.
As shown in Figure~\ref{fig:u-net}, we configure the U-Net in such a way that it
takes as input 64x64-pixel patches from the two summary images, 
and outputs a 64x64-pixel image where each pixel represents the probability that there are spiking neurons at this pixel during this time segment.
The patch-based approach is because we should be able to segment spatially distant neurons independently, 
and having a smaller input reduces the overall model size and makes the model easier to train. 

%

To train the U-Net, we synthesize training datasets by voltage imaging simulation in order to be able to generate a larger number and variety of datasets than the real voltage imaging datasets we have, without having to annotate them.
Figure~\ref{fig:train} shows an example frame from a synthesized video and a binary mask indicating the locations of spiking neurons.

We synthesize 1,000 videos with varying configurations of neurons, blood vessels, illumination, and noise.
Each video has 1,000 frames of 128x128 pixels, and after motion correction and summarization, we have 20 summary image pairs.
For each summary image pair, we randomly pick 10 of 64x64-pixel patches (overlaps are allowed), resulting in 200 patches.
Hence, in total we feed 200,000 patches to the U-Net for training, where 20\% of them are used for validation.
We use the binary cross-entropy loss and RMSProp optimizer.


\begin{figure}[t]
    \centering
    \includegraphics[trim=0 300 40 0,clip,width=\columnwidth]{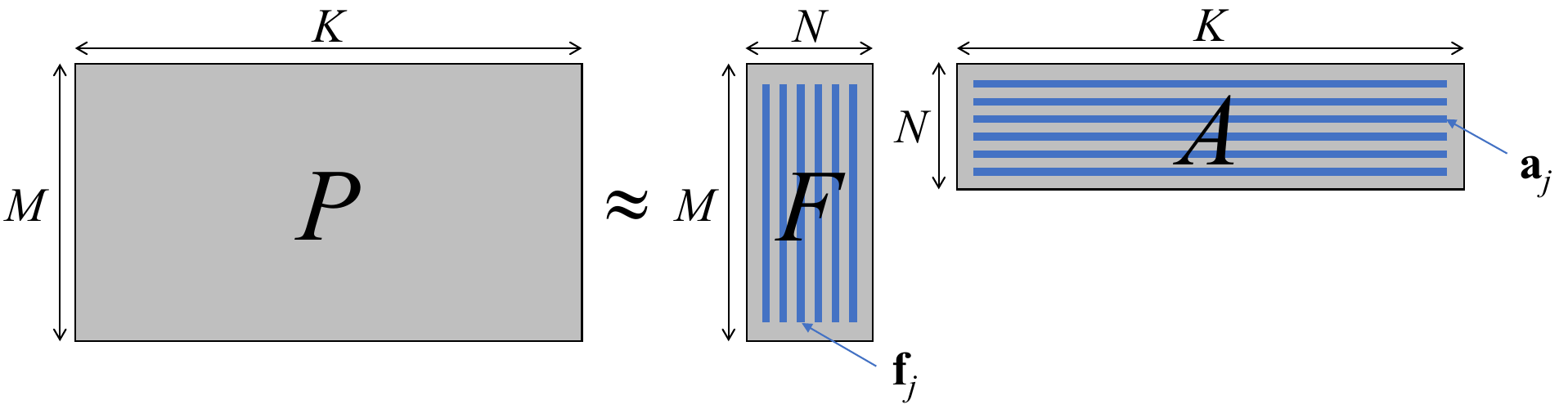}
    \caption{NMF decomposition of a sequence of probability maps, represented as a matrix $P$, into neuron footprints $F$ and their temporal activities $A$.}
    \label{fig:nmf}
\vspace{5mm}
    \centering
    \includegraphics[width=0.4\columnwidth]{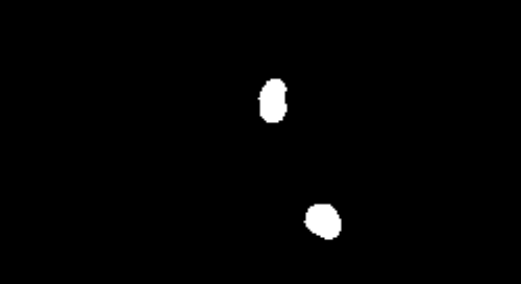}
    \includegraphics[width=0.4\columnwidth]{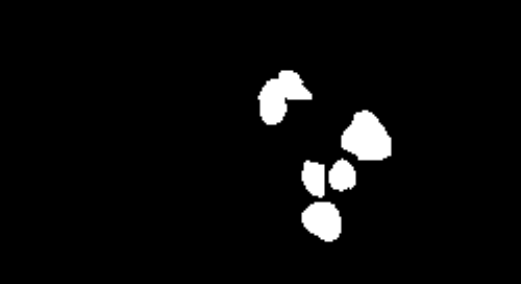}\\
    \mbox{\footnotesize (a) Individual mask $b_i(x)$ \hspace{5mm} (b) Aggregated mask $b(x)$}
    \caption{Binary masks showing areas that potentially contain neuron footprints.
    (a) Mask from one time segment, corresponding to the left image of Figure~\ref{fig:u-net_outputs}.
    (b) Mask aggregating those from all of the time segments.}
    \label{fig:binary_masks}
\end{figure}
    
We apply the trained U-Net to test data by taking sliding patches as shown in Figure~\ref{fig:segmentation},
and merge U-Net outputs to reconstruct a single probability map for each summary image pair via weighted average.
Output probability maps corresponding to the temporal summary inputs in Figure~\ref{fig:max-median_images} are shown in Figure~\ref{fig:u-net_outputs}.
Although some small blobs are incorrectly extracted, most noise and blood vessels are suppressed, and the locations of spiking neurons are delineated.

\subsection{Neuron Footprint Reconstruction}

Given a sequence of probability maps $p_i(x)$ indicating the likelihood of spiking neurons at pixel $x$ during the $i$-th time segment, we decompose it into $N$ neuron footprints and their temporal activity profiles by NMF as (see Figure~\ref{fig:nmf}):
\begin{equation}
\label{eqn:nmf}
P \approx F A.
\end{equation}
Here, $p_i(x)$ is treated as a matrix $P \in \mathbb{R}^{M \times K}$, where $M$ is the number of pixels per video frame and $K$ is the number of time segments. 
The matrix
$F \in \mathbb{R}^{M \times N}$
consists of $N$ column vectors $\mathbf{f}_j \in \mathbb{R}^M$ representing the $j$-th neuron footprint, and the matrix
$A \in \mathbb{R}^{N \times K}$
consists of $N$ row vectors $\mathbf{a}_j \in \mathbb{R}^K$ representing the temporal profile of the $j$-th neuron.

In reality, applying NMF in Equation~\ref{eqn:nmf} directly 
does not produce good results because the U-Net outputs have some spurious detections, which can translate to false components.
Moreover, NMF requires the number $N$ of neurons as input, which is challenging to estimate.
Therefore, we find the following procedure to be more reliable.
First, we threshold the probability maps $p_i(x)$ 
to obtain binary masks $b_i(x)$, 
eliminating U-Net detections with small probabilities.
We further eliminate regions whose shape is unlikely to be due to neurons by looking at their area, concaveness, and eccentricity.
An example result of this process applied to the left image of Figure~\ref{fig:u-net_outputs} is shown in Figure~\ref{fig:binary_masks}(a), where small blobs have been removed.
After that, we project these cleaned-up masks along time segments by logically OR-ing them to obtain a single binary mask as $b(x) = \bigvee_{i=1}^K b_i(x)$, which indicates potential areas of neuron footprints as shown in Figure~\ref{fig:binary_masks}(b).
Then, each connected component of $b(x)$ is where a few neurons might overlap (as visually noticeable at the top of Figure~\ref{fig:binary_masks}(b)), to which we apply NMF individually.
By confining NMF to a small area, we can keep the matrix size as well as the potential number $N$ of neurons small, making factorization more stable and faster.
%
The bottom left image of Figure~\ref{fig:pipeline} shows reconstructed neuron footprints. 

\begin{figure*}[t]
    \centering
    \includegraphics[trim=8 26 4 51,clip,width=\linewidth]{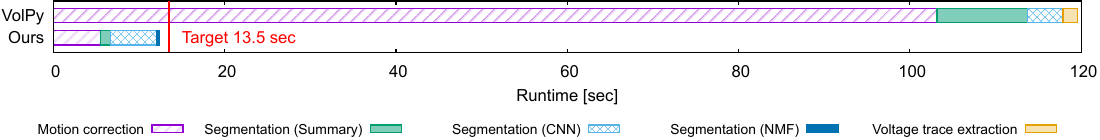}
    \caption{Runtimes of VolPy \cite{VolPy} and our pipeline for one of HPC2 datasets shown in Figure~\ref{fig:pipeline}.
    A shorter runtime is better.
    Our target is to process it within the video recording time of 13.5 sec.
    Our pipeline meets the target.}
    \label{fig:02_08_runtime}
\end{figure*}

\section{Evaluation}
\label{sec:eval}

We run our pipeline\footnote{\markchanges{Available at \url{https://github.com/mitmedialab/voltage}}} 
%
%
on a single desktop computer 
as in Table~\ref{tab:computer}
using two GPUs for motion correction and the U-Net.

\begin{table}[h]
\caption{Computational Environment}
\begin{center}
\begin{tabular}{|c|c|}
\hline
\textbf{Component} & \textbf{Specification} \\
\hline
CPU & AMD Ryzen Threadripper 3960X (24 cores) \\
RAM & DDR4 3200 MHz, 192 GB (6 ch. $\times$ 32 GB) \\
GPU & 2 of NVIDIA GeForce RTX 2080 Ti \\
SSD & KIOXIA EXCERIA PRO 2 TB (PCIe Gen 4 ${\times}4$) \\
OS  & Ubuntu 20.04.6 LTS, Linux kernel 5.15 \\
\multirow{2}{*}{Software}
    & NVIDIA Driver 525.105.17, CUDA 12.0 \\
    & Python 3.8, TensorFlow 2.4.1 \\
\hline
\end{tabular}
\label{tab:computer}
\end{center}
\end{table}


Table~\ref{tab:datasets} shows datasets we use. 
Each dataset group includes from 3 to 13 videos, totaling 37 videos.
Each video has 10,000 through 20,000 frames. 
The first three dataset groups are curated and annotated by the VolPy authors \cite{VolPyData}. 
Here we introduce a new dataset group\footnote{\markchanges{Available at \url{https://zenodo.org/records/10020273}}},
named ``HPC2,'' consisting of 13 videos capturing mouse hippocampi using SomArchon voltage indicator \cite{SomArchon}.
While HPC uses patterned illumination to 
reduce background clutter \cite{paQuasAr}, HPC2 uses normal one-photon wide-field microscopy. 
Hence, we believe HPC2 to be a useful addition as conventional microscopy data of deep-brain voltage imaging. 

\begin{table}[h]
\caption{Datasets}
\begin{center}
\begin{tabular}{|c|c|c|c|c|}
\hline
\textbf{Dataset} & \textbf{Animal \&}    & \textbf{Frame}      & \textbf{Voltage} \\
\textbf{group}   & \textbf{brain region} & \textbf{rate (fps)} & \textbf{indicator} \\
\hline
L1   \cite{VolPyData}  & Mouse      L1 cortex    &   400   & Voltron     \cite{Voltron}     \\
TEG  \cite{VolPyData}  & Zebrafish  Tegmentum    &   300   & Voltron     \cite{Voltron}     \\
HPC  \cite{VolPyData}  & Mouse      Hippocampus  & 1,000   & paQuasAr3-s \cite{paQuasAr}    \\
HPC2               & Mouse      Hippocampus  & 645-826 & SomArchon   \cite{SomArchon}   \\
\hline
\end{tabular}
\label{tab:datasets}
\end{center}
\end{table}

\subsection{Speed Evaluation}
\label{sec:speed}

\begin{figure}[t]
    \centering
    \includegraphics[trim=15 35 15 42,clip,width=\columnwidth]{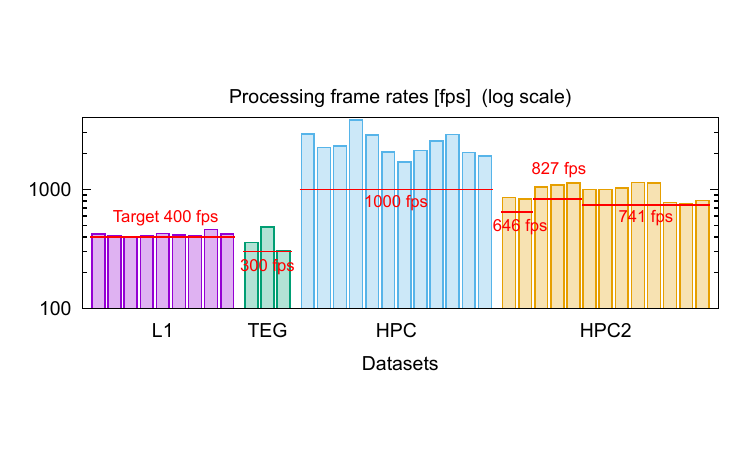}
    \caption{Processing speeds of our pipeline in frame rates for individual datasets.
    Higher rates are better.
    Our target is to process a video at its recording frame rate.
    Our pipeline meets the target for all the datasets.}
    \label{fig:frame_rates}
\vspace{5mm}
    \centering
    \includegraphics[trim=15 35 15 42,clip,width=\columnwidth]{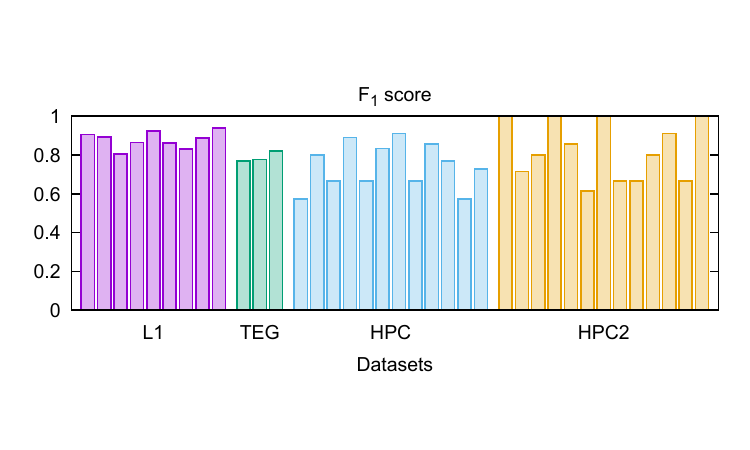}
    \caption{$F_1$ scores of our neuron segmentation method for individual datasets.
    The possible range of scores is [0, 1].
    Higher scores are better.}
    \label{fig:scores}
\end{figure}

We begin by reporting the speed of our pipeline and that of VolPy 
for the example dataset from HPC2 shown in Figure~\ref{fig:pipeline}.
Figure~\ref{fig:02_08_runtime} plots the runtimes of individual stages of each pipeline 
as well as a breakdown of the segmentation stage of each method into summary image generation, CNN (either U-Net or Mask R-CNN), and NMF (only used by our pipeline).
Our target processing time is the recording time of this video, which is 13.5 sec.
Our pipeline finishes processing within the target time, 
while VolPy's runtime is 8.9 times longer. 

\markchanges{
For all of the datasets, Volpy's runtime is longer than the video recording time
(6.4, 7.8, 4.2, and 7.2 times longer on average for L1, TEG, HPC, and HPC2, respectively)},
whereas our pipeline achieves real-time processing. 
Figure~\ref{fig:frame_rates} shows the processing speeds of our pipeline for individual datasets expressed in frame rates.
For each dataset, its video recording frame rate is set as a target indicated by the horizontal bars.
The processing speeds exceed the respective targets for all of the datasets, demonstrating real-time processing speeds.

\subsection{Accuracy Evaluation}

In order to assess segmentation accuracy, we rely on human annotation.
While the agreement with human annotation by no means signifies whether a given method correctly detects true neurons, it represents how well it can replace manual labor that is routinely done in research \cite{Voltron,SomArchon,Simul2pVoltage,HighSpeed2pVoltage}.
We follow previous work \cite{CaImAn,VolPy} and use the $F_1$ score as an accuracy metric
at an intersection-over-union threshold of 0.3.

\begin{figure}[t]
\centering
\includegraphics[trim=0 125 585 0,clip,width=0.79\columnwidth]{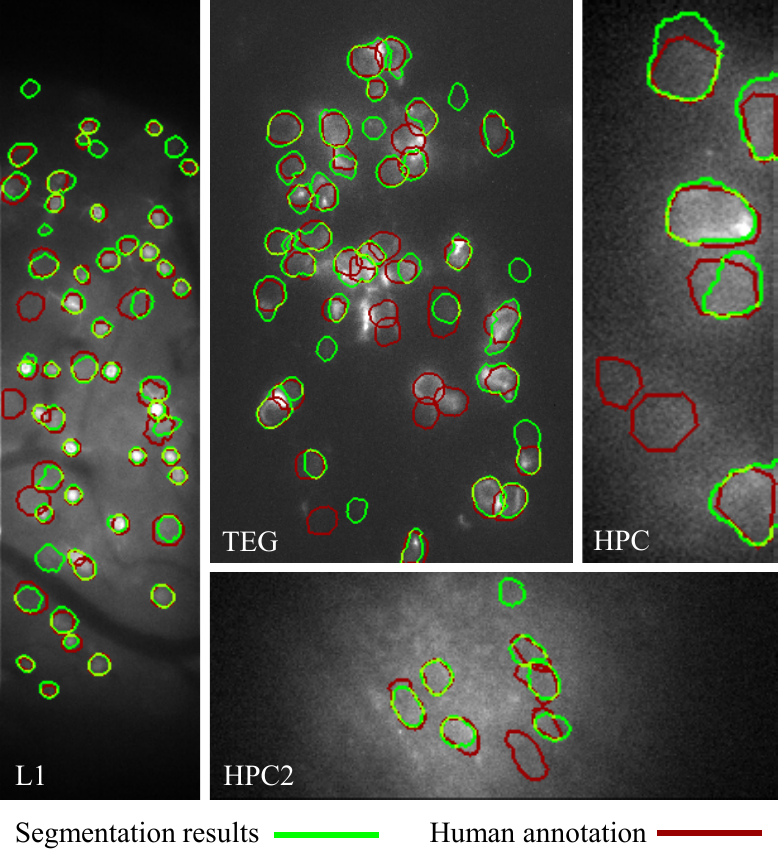}
\caption{Example results of our segmentation method.}
\label{fig:examples}
\end{figure}

Figure~\ref{fig:scores} shows the $F_1$ scores of our segmentation method for individual datasets.
The scores range from around 0.6 to 1.0, and our method achieves an $F_1$ score of 0.8 on average.
Figure~\ref{fig:examples} shows some of the segmentation results.

Table~\ref{tab:accuracy} shows the average $F_1$ scores \markchanges{(along with precision and recall)} of VolPy and our method for each dataset group.
For L1, TEG, and HPC, the VolPy scores are taken from their paper \cite{VolPy}.
%
For HPC2, 
we evaluated VolPy through leave-one-out cross-validation.
Namely, each one of the 13 videos was processed by the Mask R-CNN trained on the remaining 12 videos.
Note that our U-Net was trained on synthesized data alone without using any of the test datasets.
%
Table~\ref{tab:accuracy} indicates that VolPy and our method perform roughly equally well for cleaner datasets L1 and TEG.
In contrast, HPC has more noise and background fluorescence even with patterned illumination, and HPC2 has even more cluttered backgrounds.
Our method maintains high $F_1$ scores for these datasets.

\begin{table}[h]
    \caption{Per-group average segmentation accuracy}
    \begin{center}
    \begin{tabular}{|c|ccc|ccc|}
    \hline
    \textbf{Dataset} & \multicolumn{3}{c|}{\textbf{VolPy}} & \multicolumn{3}{c|}{\textbf{Ours}} \\
    \textbf{group} & \markchanges{Prec.} & \markchanges{Recall} & $F_1$ & \markchanges{Prec.} & \markchanges{Recall} & $F_1$ \\
    \hline 
    L1   & {\bf 0.92} & {\bf 0.88} & {\bf 0.90} &      0.91  &      0.85  &      0.88   \\
    TEG  &      0.78  &     0.74   &      0.76  & {\bf 0.83} & {\bf 0.76} & {\bf 0.79}  \\
    HPC  &      0.61  &     0.77   &      0.66  & {\bf 0.75} &      0.77  & {\bf 0.74}  \\
    HPC2 &      0.42  &     0.51   &      0.38  & {\bf 0.89} & {\bf 0.79} & {\bf 0.82}  \\
    \hline
    \end{tabular}
    \label{tab:accuracy}
    \end{center}
\end{table}

\section{Conclusion}


We have proposed a fast neuron segmentation method for voltage imaging,
and demonstrated real-time processing on a single desktop computer for the first time.
Future work includes experimenting with newer computer hardware and other CNN architectures,
as well as incorporating better voltage trace extraction methods while accelerating them.




%

\bibliographystyle{IEEEtran}
\bibliography{IEEEabrv,voltage_bibm}

\end{document}